\documentclass[5p,twocolumn]{revtex4-2} 
\usepackage{graphicx}
\usepackage[table]{xcolor}
\usepackage{amsmath}
\usepackage{amssymb}
\usepackage{epstopdf}
\usepackage{multirow}

\begin{document}

\oddsidemargin = -12pt
\title{The spectrum of $^{18}$O in terms of the multiconfigurational dynamical symmetry}


\author{G. Riczu $^{1}$}%
 \email{riczugabor@atomki.hu}
\author{C.S. Panda $^{1, 2}$}
 \email{chandra.sekhar.panda@atomki.hu}
\author{G. Lévai $^{1}$}%
 \email{levai@atomki.hu}
\author{J. Cseh $^{1}$}
 \email{cseh@atomki.hu}
 \affiliation{%
$^{1}$ HUN-REN Institute for Nuclear Research, P. O. Box 51, 4001 Debrecen, Hungary\\
$^{2}$ Doctoral School of Physics, University of Debrecen, 4026 Debrecen, Bem tér 18/B, Hungary\\
}%


\date{\today}

\begin{abstract}
\noindent
\textbf{Background:} The experimental study of the $^{6}$Li-transfer reaction and alpha-scattering resulted in a rich spectrum of $^{18}$O, including some strong indication for band structure. Different  configurations are proposed for the structural interpretation of different energy domains.  
\\
\textbf{Purpose:} We intend to investigate whether a uniform description is possible for the complete spectrum, including the high-lying states from scattering measurement. 
\\
\textbf{Methods:} 
We apply the multiconfigurational dynamical symmetry for the determination of the shape isomers and for the calculation of the spectrum obtained from different reactions. 
\\
\textbf{Results:} 
The shape isomers of $^{18}$O is obtained from the study of the stability and self-consistency of the SU(3) symmetry, which is in the intersection of the shell and different cluster configurations.
The  $^{14}$C+$^{4}$He spectrum is calculated, and its connection to other configurations is revealed. A very simple Hamiltonian is applied, including only a harmonic oscillator term, a quadrupole interaction and an $L^2$ term.
\\ 
\textbf{Conclusions:} 
The spectrum  (from the ground state up to the high-lying states) 
can be reasonably well described  by a   $^{14}$C+$^{4}$He cluster configuration.
 All the state allows the $^{12}$C+$2n$+$^{4}$He clusterization as well, and has definite shell model content. For the lowest-lying three bands the shell configuration is unique, and the overlap of the shell and the different cluster configurations is 100\%.
\end{abstract}

\maketitle


\maketitle

\section{Introduction}
The competition or coexistence of cluster and shell model degrees of freedom is an evergreen topic for the structure investigation of atomic nuclei. For a long time this kind of studies concentrated mainly on alpha-like ($N=Z=$even) nuclei, see e.g. 
\cite{ptpsupl68}.
More recently, however, much attention is focused on systems with extra neutrons
\cite{extran}.
In particular, the concept of covalent neutron configurations have been invented, and investigated extensively
\cite{covalentn}.

The $^{18}$O nucleus seems to be an especially interesting example. 
Looking at it from the cluster aspect it is remarkable that the centers of gravity of the mass and charge distributions do not coincide. 
Furthermore, according to the arguments of
\cite{vonOertzen}
the $^{14}$C nucleus can be considered as an especially good cluster, similar to
the $^{16}$O, since it has similar properties.
In particular, the first excited state of $^{14}$C
is relatively high-lying (6 MeV), and it has large proton (20.83 MeV), neutron (8.17 MeV), and alpha-binding energy (12.012 MeV), too. 

Several recent experiments revealed a rich spectrum of $^{18}$O, including highly excited regions.
In
\cite{vonOertzen}
e.g. a 
$^{12}$C($^{7}$Li,p)
transfer was measured, which resulted in 30 new states. Based on a  comprehensive overview of the available experimental and theoretical studies the authors established 6 rotational bands.
They were interpreted in terms of shell-like, core-plus-alpha, and 
$^{12}$C+$2n$+$\alpha$
cluster structures.
The interpretation of the structure is, however, not unequivocal. A newer measurement of the $^{12}$C($^{7}$Li,p) reaction,
which was designed especially for the determination of the alpha-widths puts some question marks on the importance of the alpha-cluster configurations
\cite{pirrie20}.

On the other hand alpha-scattering resonances have been observed in a large number in
\cite{roga1,roga2}.
Please note that populating a state in elastic alpha-scattering is an evidence in itself for the presence of a core-plus-alpha configuration in its structure (to some extent).

The shell-like and the cluster 
$^{14}$C+$\alpha$ and $^{12}$C+$2n$+$\alpha$ configurations were considered in
\cite{vonOertzen}
as distinct structures, which are separated in energy.
Their locations and association to the experimental bands are obtained intuitively, governed by binding energy argument, but their microscopic connection are not really investigated. 
It is not completely clear either, to what extent they can be described together with the resonances of the alpha scattering in a unified way.

These open questions inspired our present study. Here we describe the well-established bands together with the high-lying resonance spectrum in a framework, in which both the shell, and the cluster  configurations (both binary and with extra neutrons) are taken into account. In order to do so we apply the multiconfigurational dynamical symmetry (MUSY), which is the common intersection of the shell, collective, and cluster model for multi-shell problems
\cite{musy3, musysym, musy4}.
MUSY provides us with a unified multiplet structure of the shell and different cluster configurations. Furthermore, we apply a Hamiltonian, which is invariant under the transformations between the various configurations. 
We pay special attention to the interrelation of the different configurations proposed previously.
Considering the strong cluster-like features of the $^{14}$C nucleus, we focus on the  spectrum of the $^{14}$C+$\alpha$ configuration, which allows also  $^{14}$C+$2n$+$\alpha$ clusterization, and has well-defined shell-model content.
In some cases the wave functions of these configurations have 100\% overlap. In general, we determine  the (relative) alpha-cluster spectroscopic factor of each state in a  microscopic way. 
  
In what follows first we review some previous investigations in Section II, then the basic features of the MUSY are presented in Section III. 
The calculation of the model spaces is described in Section IV. We investigate the shape isomers of 
the $^{18}$O nucleus, and their clusterizations in Section V. The spectrum (energy, electromagnetic transitions, spectroscopic factors) is presented in Section VI.
Finally a brief summary is given and some conclusions are drawn in Section VII.

\section{Background}
Much experimental information was gathered  in the new millennium on the $^{18}$O nucleus, due to the investigation of two reactions. 
The $^{12}$C($^{7}$Li,p) transfer was measured first in 
\cite{vonOertzen}, then in
\cite{pirrie20}.
These studies resulted in many new states, a well-established band structure, and alpha spectroscopic factors. In particular, the $K^{\pi}$ = $0^{+}_1, 1^{-}_1, 0^{+}_2, 0^{-}_2, 0^{+}_4, 0^{-}_4$ bands of Figure 4 were constructed in 
\cite{vonOertzen}, based on the new and previous data. The first two bands were considered to have shell-like structure, while the other four are considered to be parity-doublets, and have cluster structure of alpha-particle plus $^{14}$C, or ($^{12}$C + 2$n$).   

The
\cite{pirrie20}
experiment was performed in order to determine branching ratios for states in $^{18}$O.
The $\Gamma_{\alpha} / \Gamma_{tot}$ values were measured, and reduced ${\alpha}$-widths were compared with the Wigner limit, resulting in spectroscopic factors. Some states displayed cluster structure in this investigation, but no evidence of consistent cluster structure is seen across the bands previously proposed.

The other reaction studied in detail was the alpha-scattering 
\cite{roga1,roga2}.
In
\cite{roga2} 
the $^{14}$C+$\alpha$ elastic scattering was studied with the thick target inverse kinematic techniques. The data were analyzed with the multichannel multilevel R-matrix theory, and resonance data were obtained in the excitation energy range of 8-15 MeV. Alpha spectroscopic factors ($\theta_{\alpha}$)  are given as the ratio of the alpha reduced width and the single particle value.
Broad, purely $\alpha$-cluster $0^+$ and $2^+$ states at 9.9 and 12.9 MeV were observed. On the other hand
from the observed characteristics of the states the authors do not see evidence for the inversion doublet bands, suggested by von Oertzen et al. The splitting of alpha-cluster states is conjectured. 

It is interesting to see that the alpha reduced widths were obtained form the transfer and scattering experiments in different ways 
(i.e. by reconstruction of decay products, and R-matrix analysis of the cross section).
A careful comparison of the two methods could be of interest.

As for the earlier experimental studies we refer to the reviews of these recent papers.

From the theoretical side different approaches have been applied to the $^{18}$O nucleus and its possible alpha-cluster structure.  The generator coordinate method (GCM) calculation
\cite{gcm}
as well as the coupled channel orthogonality condition model
\cite{ocm}
predicted alpha-cluster bands.
A phenomenological algebraic model gave a molecular band with alternating parity
\cite{alternating}.
In a more recent antisymmetrized molecular dynamics calculation, which was combined with GCM
\cite{furutachi}, 
it was found that the  
$K^{\pi} = 0^{+}_{2}$ and $0^{-}$ rotational bands have prominent core-plus-alpha cluster structure.
It is also noted that the $^{14}$C+$\alpha$ cluster structure is fragmented into many states. 

The semimicroscopic algebraic cluster model (SACM) was also applied to the $^{18}$O
nucleus in a previous study
\cite{levai92}.
Our present work has several features in common with that previous study.  Nevertheless, there are also considerable differences, as follows.  (i) Now we have a much richer experimental data, with better-defined band-structure. (ii) In    
\cite{levai92}
only the $^{14}$C+$\alpha$ cluster structure was taken into account, while here we consider also the shell-structure, and the $^{12}$C+$2n$+$\alpha$ configuration, and their interrelations. In particular, we apply interactions, which are invariant with respect to the transformations from one configuration to the other. (iii) We use a Hamiltonian, which has only three parameters, as opposed to the seven parameters in
\cite{levai92}. (iv) The phenomenological parameters of our energy functional have well-defined and generally accepted physical content: the oscillator energy, the strength of the quadrupole force, and the $L^2$ interaction. (iv) We consider some new aspects of the problem, like the presence of shape isomers, and the alpha spectroscopic factors.

\section{Multiconfigurational dynamical symmetry}
	A unified systematization of the shell model, collective model, and cluster model is found in the multi-configurational dynamical symmetry  (MUSY) \cite{musy3, musysym, musy4}. The MUSY has the ability to connect the wave functions and energy spectra of different configurations. This represents a multi-shell extension of the historical relationship established in 1958 for a single-shell problem \cite{elliott, K. Wildermuth, baybohr}. It constitutes a composite symmetry where each  configuration possesses a usual [U(3)] dynamical symmetry, and an additional symmetry which connects these configurations among themselves.\\
For single-major-shell problem, the shell or quartet configuration is defined by the algebra chain: 
\begin{equation}
	\begin{aligned}
	 U(3)\:\supset\;\; SU(3) \;\supset \quad SO(3)\\
		|[n_{1}, n_{2}, n_{3}],\;(\lambda, \mu),\; K, \;\quad L\quad \rangle.
	\end{aligned}
\end{equation}
Here $\lambda$=$n_{1}-n_{2}$, $\mu$=$n_{2}-n_{3}$ and $K$ is the multiplicity of the orbital momentum $L$. Only the space symmetry of the states is indicated here. The spin-isospin part is characterized by Wigner's $U^{ST}(4)$ group \cite{E.P. Wigner}.  Any cluster or collective wave function can be simplified by expanding it in terms of the U(3) basis in the shell model. The U(3) group can be obtained from U(N) groups, which share the same Young pattern with the permutation group. Here, N represents the number of orbitals in a single major shell of the harmonic oscillator. The many-body wave function's anti-symmetrization is satisfied by the associate representation of U(N) and $U^{ST}(4)$, where their Young patterns are related by reflection: interchanging the rows and columns \cite{symminphy}. For problems involving multiple major shells, this procedure is iteratively repeated.\\
For the binary cluster configuration, the space part is defined by the group structure:
\begin{equation}
	\begin{split}
		U_{C_{1}}(3)\otimes U_{C_{2}}(3)\otimes U_{R}(3)\supset U_{C}(3)\otimes U_{R}(3)\supset U(3)\\
		\supset SU(3)\supset SO(3)
	\end{split}
\end{equation}
Here, $C_{1}$ and $C_{2}$ refer to cluster numbers 1 and 2, and $U_{C}(3)$ stands for the coupled space symmetry of the two clusters. $U_{R}(3)$ represents the relative motion. The Elliott model \cite{elliott} defines the internal structure of the cluster, whereas the modified $[U_{R}(4)]$ vibron model \cite{F. Iachello} accounts for the relative motion of the cluster parts. (Here R stands for the relative motion of clusters.)

A state of an $[n_{1},n_{2},n_{3}]$ symmetry exists in the binary cluster configuration if the triple product matches with it:
\begin{equation}
	\begin{split}
		[n_{1}^{C_{1}},n_{2}^{C_{1}},n_{3}^{C_{1}}]\otimes[n_{1}^{C_{2}},n_{2}^{C_{2}},n_{3}^{C_{2}}]\otimes[n_{R}, 0, 0]=\\
		[n_{1},n_{2},n_{3}]+...
	\end{split}
\end{equation}
The value of $n_{R}$ is limited to the lowest value due to the Pauli principle (known as the Wildermuth condition \cite{K. Wildermuth}).
The application of a Hamiltonian featuring U(3) dynamical symmetry allows for the analytical calculation of energy in a unified manner across different configurations. The same approach is applicable for electromagnetic transitions.\\
For the many-major-shell problem, a unified classification scheme is defined by the algebraic chain:
\begin{equation}
	\begin{aligned}
		U_s(3) \;\;\quad \otimes\;\;\quad U_e(3)\; \supset\quad U(3)\:\supset\;\; SU(3) \;\supset \quad SO(3)\\
		|[n_{1}^{s}, n_{2}^{s}, n_{3}^{s}], [n_{1}^{e}, n_{2}^{e}, n_{3}^{e}], \rho, [n_{1}, n_{2}, n_{3}],\;\;(\lambda, \mu), K, \quad L\quad \rangle.
	\end{aligned}
\end{equation}
encompassing the shell, collective, and cluster models. This symmetry is the common intersection of the symplectic shell model \cite{sympl1, sympl2}, the contracted symplectic model \cite{contrsympl1, contrsympl2}, and the semi-microscopic algebraic cluster model \cite {sacm1, sacm2}. For the shell and collective models $s$ stands for the band head; for cluster it refers to the internal cluster structer. $e$ indicates in each case the major shell excitations; in the shell and collective model case it takes place in steps of 2$\hbar\omega$, connecting oscillator shells of same parity, while in the cluster case it is in steps of 1$\hbar\omega$, incorporating all the major shells. For the cluster model it has only completely symmetric (Single row Young-tableaux) irreducible representations (irreps): [n, 0, 0], while in the case of the shell and collective models it can be more general. As a consequence the model spaces of the three models have a considerable overlap, but they are not identical.\\
The MUSY can be obtained by selecting a Hamiltonian that remains unchanged under transformations in the pseudo-space of particle-indices \cite{musy3, musysym, musy4}. This can be achieved by expressing the Hamiltonian in terms of the operators of the second part of chain (4) then the operators are invariant with respect to the transformations from one configuration to the other. A simple Hamiltonian of this kind can be written using the Casimir operators of the algebras $U (3)\supset SU (3) \supset SO(3)$:
\begin{equation}
	{\hat H} = (\hbar \omega) {\hat n} + a{\hat C}_{SU(3)}^{(2)} + \frac{d}{2\theta}{\hat L}^2.
\end{equation}

The first term $\lbrace(\hbar\omega){\hat n}\rbrace$ represents the harmonic oscillator Hamiltonian (a linear invariant of the U(3)). The second-order Casimir operator of $SU(3)$, denoted as ${\hat C}_{SU(3)}^{(2)}$, describes the quadruple-quadruple interaction. The eigenvalues of ${\hat C}_{SU(3)}^{(2)}$ for an SU(3) basis quantum number $(\lambda, \mu)$ are given by $\lambda^{2} + \mu^{2} + \lambda\mu + 3(\lambda + \mu)$. $\theta$ is the moment of inertia, calculated for an ellipsoid (with cylindrical symmetry). Please note that this Hamiltonian is similar to that of the Elliott model
		\cite{elliott}
		the only difference is that (i) the weights of the quadrupole interaction and the rotational terms are not fixed, and (ii) the moment of inertia is not uniform for all the  bands.
		$\theta$ is calculated for a cylindrically symmetric ellipsoid. If the nucleus is triaxial, let the semi-axis lengths of the triaxial nucleus are $a$, $b$, $c$, then one can make the following approximation for a cylindrically symmetric shape with semi-axis lengths $A$, $B$, $C$: if $a/b \ge b/c$, then $A=a$ and $B=C=\sqrt{bc}$; otherwise, $A=B=\sqrt{ab}$ and $C=c$ \cite{omegas}. In this case, the moment of inertia is
\begin{equation}
	\theta  =\frac{1}{5}m(A^{2}+C^{2}).
\end{equation}
Its unit is $\hbar^2$/MeV.

\section{Model spaces}
First we construct the no-core shell model space in the $L-S$ coupled SU(3) scheme.  In this formalism the space part of the wave function is characterized by the symmetries of the
\begin{equation}
	\begin{aligned}
		U_s(3) \;\;\quad \otimes\;\;\quad U_e(3)\; \supset\quad U(3)\:\supset\;\; SU(3) \;\supset \quad SO(3)\\
		|[n_{1}^{s}, n_{2}^{s}, n_{3}^{s}], [n_{1}^{e}, n_{2}^{e}, n_{3}^{e}], \rho, [n_{1}, n_{2}, n_{3}],\;\;(\lambda, \mu), K, \quad L\quad \rangle.
	\end{aligned}
\end{equation}
group chain.
The spin-isospin part carries the symmetries of Wigner's supermultiplet theory
\begin{equation}
	\begin{aligned}
		U^{ST}(4)\quad\quad  \supset \quad SU^{S}(2)\quad \otimes\quad SU^{T}(2)\\
		|[f_{1}, f_{2}, f_{3}, f_{4} ],\quad \quad \quad \quad  S,\quad \quad \quad  \quad \quad   T\quad \rangle.
	\end{aligned}
\end{equation}
The details of the algorithm is presented in 
\cite{musy3}.
The relevant basis states up to 7 $\hbar \omega$ excitation are listed in Table I. In addition to the U(3) and SU(3) quantum numbers, also the multiplicity of the irreducible representations (irreps) is indicated as well as the expectation value of the second order Casimir operator of SU(3).
\begin{table}
	\centering
	\caption {Model space of $^{18}$O nucleus up to 7$\hbar\omega$ excitations. Here n denotes the major shell excitation, $C^{(2)}$ gives the expectation value of the second order Casimir invariant of SU(3). The states are listed according to the decreasing expectation value of $C^{(2)}$, i.e. according to the decreasing deformation. The  $m$ values are the multiplicities of the representations in the shell ($^{18}$O) and cluster ($^{14}C+\alpha$) model spaces. Some shell model basis are left out, and it is indicated  in the last column.\label{var_table}}
	
	\begin{tabular}{ccccccc}
		\hline
		\hline
		n&U(3)&SU(3)&$C^{(2)}$&$m_{Sh.}$&$m_{Cl.}$	&left out\\
		
		\hline
		0$\hbar$$\omega$&[8,4,4]& (4,0)&28&1&1&\\
		&[6,6,4]&(0,2)&10&1&&\\	
		\hline
		1$\hbar$$\omega$&[10,4,3]&(6,1)&64&1&1&\\
		&[9,5,3]&(4,2)&46&1&&\\
		&[9,4,4]&(5,0)&40&2&1&\\
		&[8,6,3]&(2,3)&34&2&&\\
		&&&&&&4\\
		\hline
		2$\hbar$$\omega$&[12,4,2]&(8,2)&114&1&1&\\
		&[11,5,2]&(6,3)&90&1&&\\
		&[11,4,3]&(7,1)&81&3&1&\\
		&&&&&&4\\
		&[10,4,4]&(6,0)&54&9&1&\\ 
		&&&&&&8\\
		\hline
		3$\hbar$$\omega$&[13,5,1]&(8,4)&148&1&&\\
		&[13,4,2]&(9,2)&136&2&1&\\
		&&&&&&4\\
		&[12,4,3]&(8,1)&100&12&1&\\
		&&&&&&4\\
		&[11,4,4]&(7,0)&70&20&1&\\
		&&&&&&11\\
		\hline
		4$\hbar$$\omega$&[15,4,1]&(11,3)&205&1&&\\
		&&&&&&4\\
		&[14,4,2]&(10,2)&160&9&1&\\
		&&&&&&6\\
		&[13,4,3]&(9,1)&121&35&1&\\
		&&&&&&4\\
		&[12,4,4]&(8,0)&88&61&1&\\
		&&&&&&15\\
		\hline
		5$\hbar$$\omega$&[16,5,0]&(11,5)&249&1&&\\
		&&&&&&5\\
		&[15,4,2]&(11,2)&186&24&1&\\
		&&&&&&7\\
		&[14,4,3]&(10,1)&144&91&1&\\
		&&&&&&4\\
		&[13,4,4]&(9,0)&108&146&1&\\
		&&&&&&19\\
		\hline
		6$\hbar$$\omega$&[18,4,0]&(14,4)&322&1&&\\
		&&&&&&7\\
		&[16,4,2]&(12,2)&214&75&1&\\
		&&&&&&6\\
		&[15,4,3]&(11,1)&169&240&1&\\
		&&&&&&6\\
		&[14,4,4]&(10,0)&130&378&1&\\
		&&&&&&23\\
		\hline
		7$\hbar$$\omega$&[19,4,0]&(15,4)&358&2&&\\
		&&&&&&7\\
		&[17,4,2]&(13,2)&244&173&1&\\
		&&&&&&5\\
		&[16, 4, 3]&(12, 1)&196&542&1&\\
		&&&&&&7\\
		&[15, 4, 4]&(11, 0)&154&822&1&\\
		&&&&&&27\\
		\hline
		\hline
	\end{tabular}
\end{table}

The $^{14}$C+$^{4}$He cluster basis is constructed in the strong (U(3)) coupled scheme. The two clusters have [4,4,2], and [0,0,0] symmetries, respectively, and the lowest Pauli-allowed value of the relative motion quantum number is 6. Physically the strong coupling means that the ground-state rotational band ($J^{\pi} = 0^+, 2^+$) of the $^{14}$C cluster is incorporated.
The Pauli-allowed cluster space (which is also free from the spurious excitation of the center of mass) is obtained by taking the intersection with the shell model space. This cluster basis is also shown in Table I, with its multiplicity. 

As for the $^{12}$C+$2n$+$^{4}$He configuration is concerned, it is allowed in each core-plus-alpha state, since the $^{12}$C+$2n$ configuration gives the [4,4,2] symmetry of the  $^{14}$C cluster. Further U(3) symmetries, and consequently larger model space is also available, but here we restrict ourselves to this smaller one, because it turns out to be rich enough to account for the experimental data. 

The states of Table I with $m=1$ shell model multiplicity have especially remarkable features. In particular in these cases the seemingly different configurations (shell, core-plus-alpha, $^{12}$C+$2n$+$^{4}$He, ...) have identical wave functions as a result of the antisymmetrization. This can be seen by expanding the wave function of any cluster configuration in the shell basis. Since both the eigenstates with different eigenvalues, and the basis states of different U(3) irreps are orthogonal to each other, the result is, that the cluster (any cluster) wave function contains only a single term in the expansion.  This is the case for the three lowest-lying bands: with quantum numbers 0[8,4,4](4,0), 1[10,4,3](6,1), 2[12,4,2](8,2).  

The unification of the clusters is illustrated in Fig. 1. The left hand part shows Harvey's prescription for the fusion of the two fragments
\cite{harvey}. 
The   $^{14}$C cluster has in each case the [4,4,2] U(3) symmetry, i.e. the quadrupole shape of the cluster is the same. But different basis sates of this irrep are relevant for the three cases. Their Cartesian coordinates  are also shown in the figure.
The different basis states of the same irrep determine  different relative orientation with respect to the molecular axis (z). In case of the [8,4,4] state the symmetry axis of the oblate  $^{14}$C is parallel with z, in case of [12,4,2] it is perpendicular (y), while in the third case it is half-way between the z and y axes. 
The central part shows the density distribution of the clusters without antisymmetrization, and the right part is the shell model wave function, being identical with the cluster wave function after the antisymmetrization.
\begin{figure*}
	
	\includegraphics[width=12.0cm]{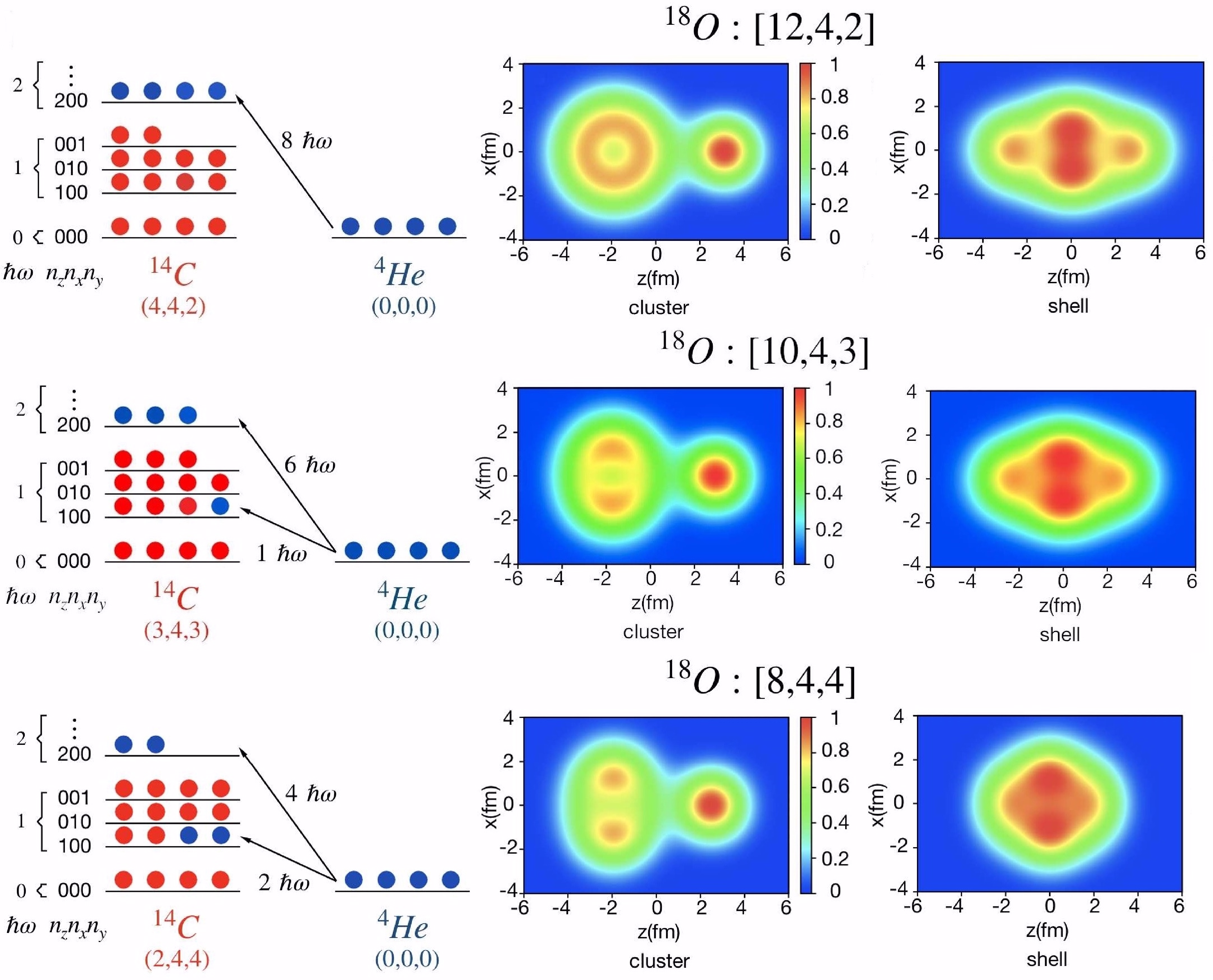}
	
	\caption{Harvey's prescription for the unification of two clusters (left). The density distributions of the clusters before the antisymmetrization is shown in the central parts, and that of the shell model is on the right side. In these cases the cluster and shell wave functions become identical as a result of the antisymmetrization.  The distances between the (centers of gravity of the) clusters are 4.436 fm, 4.691 fm, and 4.997 fm for configurations [8,4,4], [10,4,3], and [12,4,2] respectively. }
\end{figure*}

The calculation of the densities were carried out in the following way. In case of shell model: 
In the first step the length of the semi-major axis ($a,b,c$) are obtained from the self-consistency argument and the volume conservation. 
\cite{omegas}:
\begin{align}
	\frac{a}{c} &= \frac{n_{1}+\frac{A}{2}}{n_{3}+\frac{A}{2}},
	\frac{b}{c} = \frac{n_{2}+\frac{A}{2}}{n_{3}+\frac{A}{2}} \\
	c &= r_{0}\sqrt[3]{A \frac{(n_{3}+\frac{A}{2})^2}{(n_{1}+\frac{A}{2})(n_{2}+\frac{A}{2})}} .
\end{align}
Then $\omega_i$-s of the Cartesian coordinates were obtained from 
\cite{Heyde} \begin{align}
	\omega_{i} &= \omega_{0} \frac{R_{0}}{a_{i}}, \quad i=x,y,z.	
\end{align}
Where $R_{0}=\sqrt[3]{abc}$ and $r_{0}=1.2fm$, the radius parameter. $\omega_{0}$ can be determined from the relation $\hbar$$\omega_{0}$=45$A^{-1/3}-25A^{-2/3}$ \cite{Blomqvist}.

The density is calculated from the Cartesian components of the wave function:
\cite{density}.
\begin{equation}
	\rho(x, y, z, (n_{x},n_{y},n_{z}))=\sum_{i}^{A}|\psi_{i}(x, n_{x})\psi_{i}(y, n_{y})\psi_{i}(z, n_{z})|^{2}
\end{equation}
Where 
\begin{equation}
	\begin{split}
		\psi(i, n_{i})=\sqrt{\frac{1}{2^{n_{i}}n_{i}!}}\left(\frac{m \omega_{i}}{\pi\hbar}\right)^{1/4} e^{-{\frac{m\omega_{i}}{2\hbar}}i^{2}}H_{n_{i}}\left(\sqrt{\frac{m\omega_{i}}{\hbar}}i\right);\\ 
		i=x, y, z
	\end{split}
\end{equation}
is the wave function of a single nucleon along i-direction \cite{Heyde}. Here $n_{i}$ is the number of oscillator quanta and $\omega_{i}$ is the frequency along $i$-direction, and $H_{n_{i}}$ is the Hermite polynomial of order $n_{i}$. A is the total number of nucleons in the nucleus.

For the cluster model: the internal structure of the clusters are described with shell model wave functions, as above, and their relative distance is obtained from the touching configurations of the classical ellipsoids.

\section{Shape isomers}
In this section we investigate the shape isomers of the $^{18}$O nucleus. 
Their appearance is informative in general, too, but we are especially interested in the question whether or not stable shapes are present with the features as conjectured in
\cite{vonOertzen}, 
i.e. with 2 and 4 $\hbar \omega$ excitations and alpha-cluster structure.

We determine the shape isomers by applying a symmetry-governed method, called Stability and self-Consistency of the SU(3) Symmetry (SCS)
\cite{lepcsok}.
Since the  SU(3) symmetry uniquely determines the quadrupole shape of the nucleus, it is a shape-consistency and stability method
\cite{draayer,rowe}.
In particular, we perform a Nilsson-model calculation for systematic change of the
beta and gamma parameters of the quadrupole shape, and from the occupation of the Nilsson orbitals we determine the quadrupole shape of the nucleus. The output beta and gamma parameters show the stability and self-consistency of the shape. Technically the procedure is carried out by the application of the quasi-dynamical U(3) symmetry
\cite{jarrio}. 
The quasi-dynamical symmetry is a generalized version of the real (Elliott type) symmetry, which is valid even in the presence of symmetry-breaking interactions
\cite{quasidsu3}.
The shape isomers are seen in the $\beta_{in} - \beta_{out}$ plots as horizontal plateaus, as shown in Figure 2. The SCS method is an alternative to the traditional energy-minimum calculation for determining the shape isomers. Since it deals with the U(3) symmetry, it has the advantage of providing us with a selection rule for the possible clusterizations, i.e. reaction channels in which the state in question can be populated or can decay.

\begin{figure}

	\includegraphics[width=9.0cm]{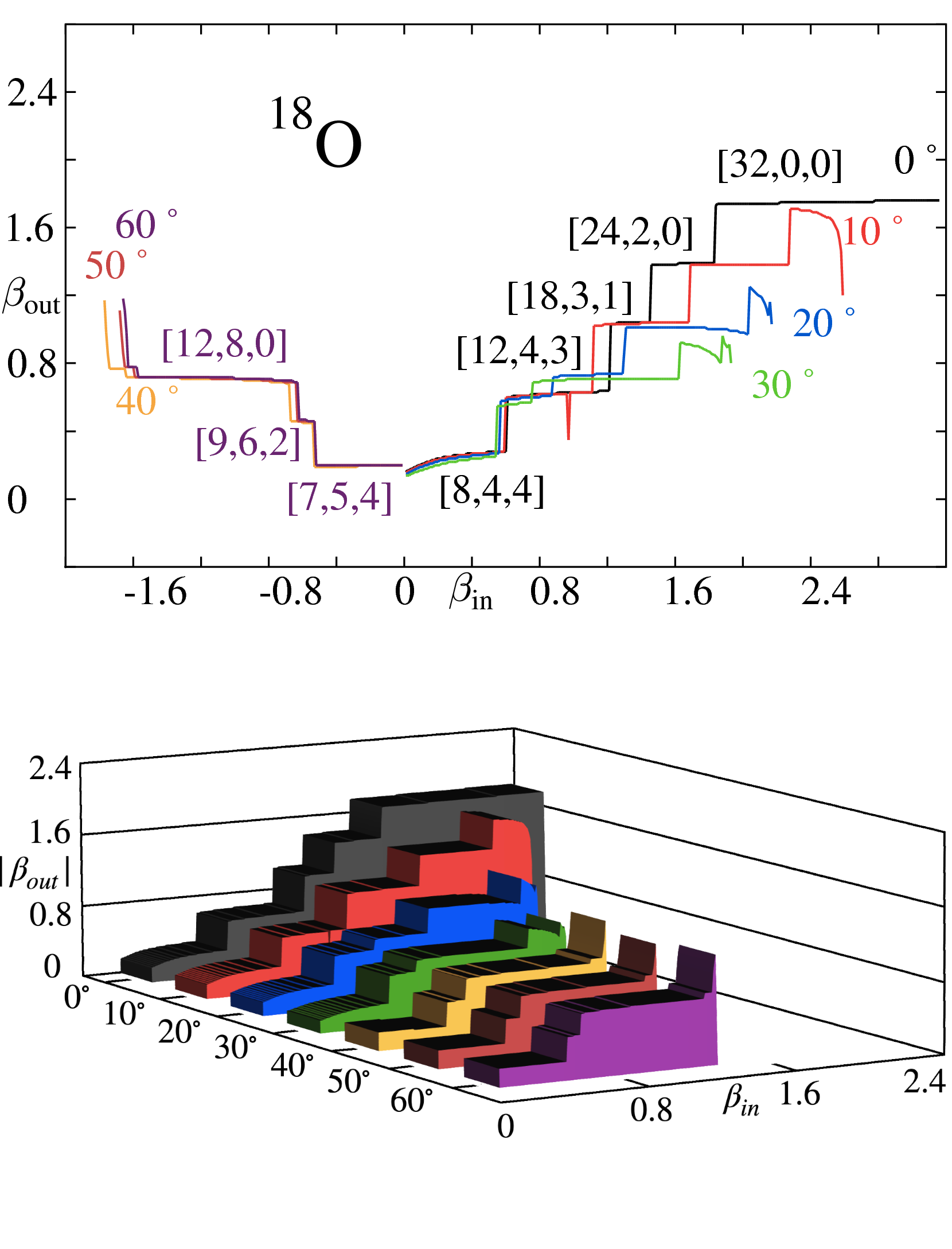}

	\caption{ Shape isomers of the $^{18}$O nucleus from the SCS method using effective U(3) quantum numbers. In the upper part, the horizontal axis shows the $\beta_{in}$ input parameter; the vertical axis indicates the value of $\beta_{out}$. The $\gamma$ parameter is given in degrees, and steps in 10. The lower part shows the same result in three dimensions}
	
\end{figure}
We have performed the Nilsson calculation with the Hamiltonian
\begin{equation}
	H = -\frac{\hbar^{2}}{2M}\Delta + \frac{M}{2}[\omega_{\perp}^{2}(x^{2} + y^{2}) + \omega_{z}^{2}z^{2}] - C({\vec{l}} \cdot {\vec{s}}\;) - D {\vec{l}}\;^{2}
\end{equation}
and the parameters of
$C$, and $D$
were taken from the systematics
\cite{nilsparam}.
The result of the calculation is shown in Figure 2, both in 2 and in 3 dimensional pictures.
The stable shapes are listed in Table II. The effective U(3) quantum numbers are obtained from the calculations, and they have some natural uncertainties
\cite{lepcsok}, 
(e.g. the resulting quantum numbers are real values, not integers
\cite{jarrio}. Therefore we indicated also a close-lying simple shell model configuration (denoted by U(3)). The ratios of the major axes of the ellipsoid, and the deformation parameters are indicated, too. 

Figure 3 shows what kind of binary clusterizations are allowed in the shape isomers. A quantity, called reciproc forbiddenness is plotted as a function of the mass number of the lighter cluster.
\begin{figure}
	\includegraphics[width=9.0cm]{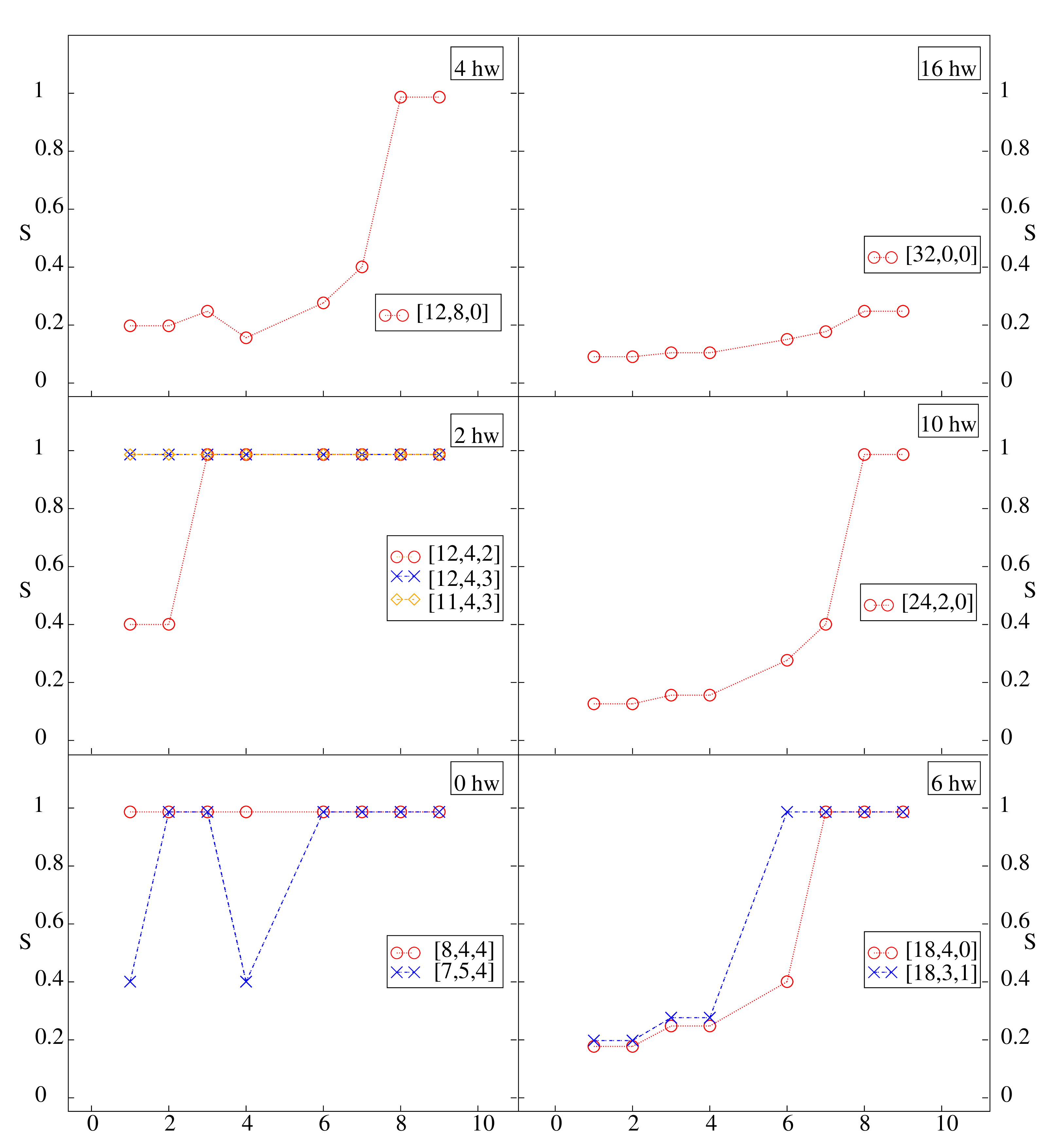}

	\caption{Reciprocal forbiddenness as a function of the mass number of the lighter cluster for the shape isomers in $^{18}$O. Those clusters are taken in which at least one cluster is stable. The lines are just to guide the eye.}
\end{figure}
If its value is 1, then the cluster configuration is structurally allowed (by the U(3) and U$^{ST}$ selection rules). Smaller values correspond to stronger forbiddenness. It is calculated as:
\begin{equation}
	S=\frac{1}{1+min(\sqrt{(\Delta n_{1})^{2}+(\Delta n_{2})^{2}+(\Delta n_{3})^{2}})},
\end{equation}
where $\Delta n_{i}=|n_{i}-n_{i,k}^{c}|$. Here $n_{i}$ refers to the U(3) representation of the parent nucleus, while $n_{i,k}^{c}$ stands for the U(3) representation of channel c, obtained from the multiplication with  quantum number of relative motion, the k index distinguishing the different product representations.

From the viewpoint of the structure interpretation of 
\cite{vonOertzen}, 
as well as regarding our forthcoming considerations the following can be said.
The 2 $\hbar \omega$ alpha-cluster state seems to correspond to the first excited shape isomer of Table II, called Pro. On the other hand, there is no 4 $\hbar \omega$ stable shape, which allows alpha-cluster structure. Therefore these states do not correspond to the ground-band of a new shape isomer (local minimum in the potential surface), rather they represent some excited states. Though there is a predicted shape isomer with 4 $\hbar \omega$ excitation, but it is a triaxial state, and does not contain alpha-cluster configuration.
\begin{table}[h]
	\centering

	\caption{ Shape isomers of the $^{18}$O nucleus. The abbreviations are as follows: GS: ground state, lin-ch: linear chain,  Pro: prolate, SD for super-deformed state, HD for hyper-deformed state. ($\beta$,$\gamma$) are the parameters of the
		quadrupole deformation ($\gamma$ is given in degrees), and a:b:c stands for the ratio of the major axes of the ellipsoid.}
	\begin{tabular}{cccccc}
		\hline
		\hline
		State&$\hbar$$\omega$&U(3)&$U_{eff}$(3)&a:b:c&($\beta$,$\gamma$)\\
		\hline
		GS&0&[8,4,4]&[8,4,4]&1.3:1.0:1&(0.31,0.0)\\
		&&&[7,5,4]&1.2:1.1:1&(0.20,19.1)\\
		\hline
		Pro&2&[12,4,2]&[12,4,2]&1.9:1.2:1&(0.67,10.9)\\
		&&&[12,4,3]&1.8:1.1:1&(0.61,5.8)\\
		&&&[11,4,3]&1.7:1.1:1&(0.55,6.6)\\
		\hline
		$Tri $&4&[12,8,0]&[12,8,0]&2.3:1.9:1&(0.74,40.9)\\
		\hline
		$SD_{pr.}$&6&[18,4,0]&[18,3,1]&2.7:1.2:1&(1.07,6.2)\\
		\hline
		HD&10&[24,2,0]&[24,2,0]&3.7:1.2:1&(1.42,4.3)\\
		\hline
		lin-ch&16&[32,0,0]&[32,0,0]&4.6:1.0:1&(1.76,0.0)\\
		\hline
		\hline
	\end{tabular}
\end{table}

A prolate superdeformed state is predicted by our SCS calculation, corresponding to 6 $\hbar \omega$ excitation, which allows the presence of alpha-cluster configuration. As it will be seen later on, some high-lying resonances of the scattering experiment may correspond to this isomer.

\section{Spectrum}
\subsection{Energy}

For the calculation of the energy we apply the dynamically symmetric Hamiltonian given in equation (5).
The parameters of the Hamiltonian are fitted to the observed data. Concerning the experimental spectrum we start with the bands established in 
\cite{vonOertzen}. 
The states with well-defined spin-parity were taken into account in the fitting procedure with a weight of 1.0, while those of uncertain spin-parities with 0.5. In addition we have investigated whether further band candidates can be found from the data of the compilation
\cite{compilation},
and from the resonances of alpha-scattering
\cite{roga2}.
In particular, if three or more states follow a straight line in the $E-J(J+1)$diagram, we considered them as a candidate band. These states were taken into account in the fitting procedure with a weight of 0.1. 
The band structure obtained in this way is shown in the lower part of Figure 4.
 	The resulting parameters values are 
	$\hbar\omega =  2.344$ MeV, $a = -0.007257$ MeV, $d = 0.8572$. 
	In Figure 5, the distribution of band-heads is shown.
		\begin{figure} [!ht]
			
			\includegraphics[width=9.0cm]{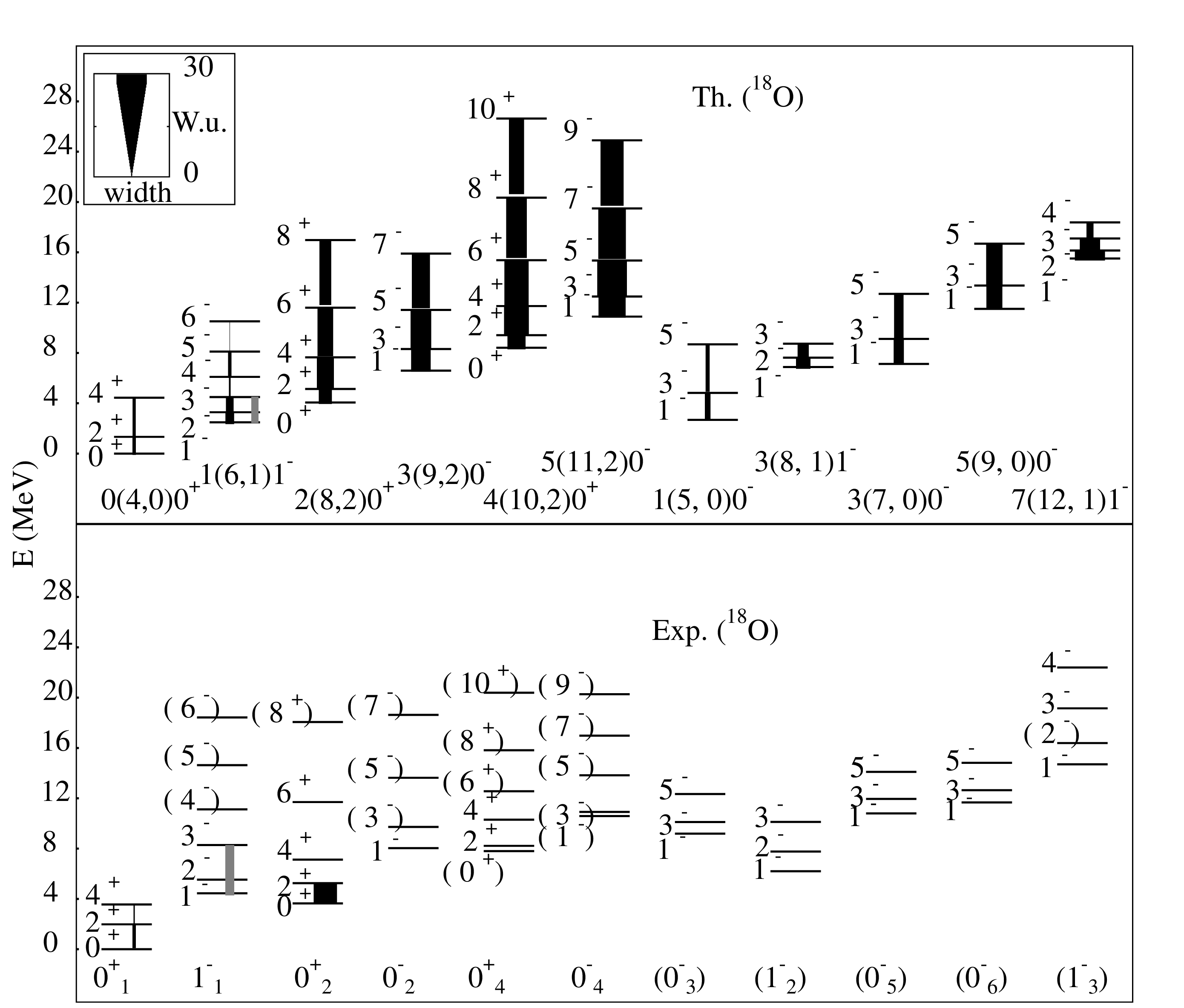}
			
			\caption{The spectrum of MUSY (upper) in  comparison with the experimental data of  $^{18}$O (lower). The bands from \cite{vonOertzen} are labeled by $K^{\pi}$ quantum numbers, while the candidate bands by ($K^{\pi}$). The model bands carry the quantum numbers $n(\lambda,\mu)K^{\pi}$. The ($0^{-}$) bands are  from  \cite{roga2} and the ($1^{-}$) ones are from the energy compilation of \cite{compilation}.   The width of the arrow between the states is proportional to the strength of the E2 transition. The gray arrow refers to transition $3^-\rightarrow1^-$. }
		\end{figure}
			\begin{figure}[!ht]
			
			\includegraphics[width=5.0cm]{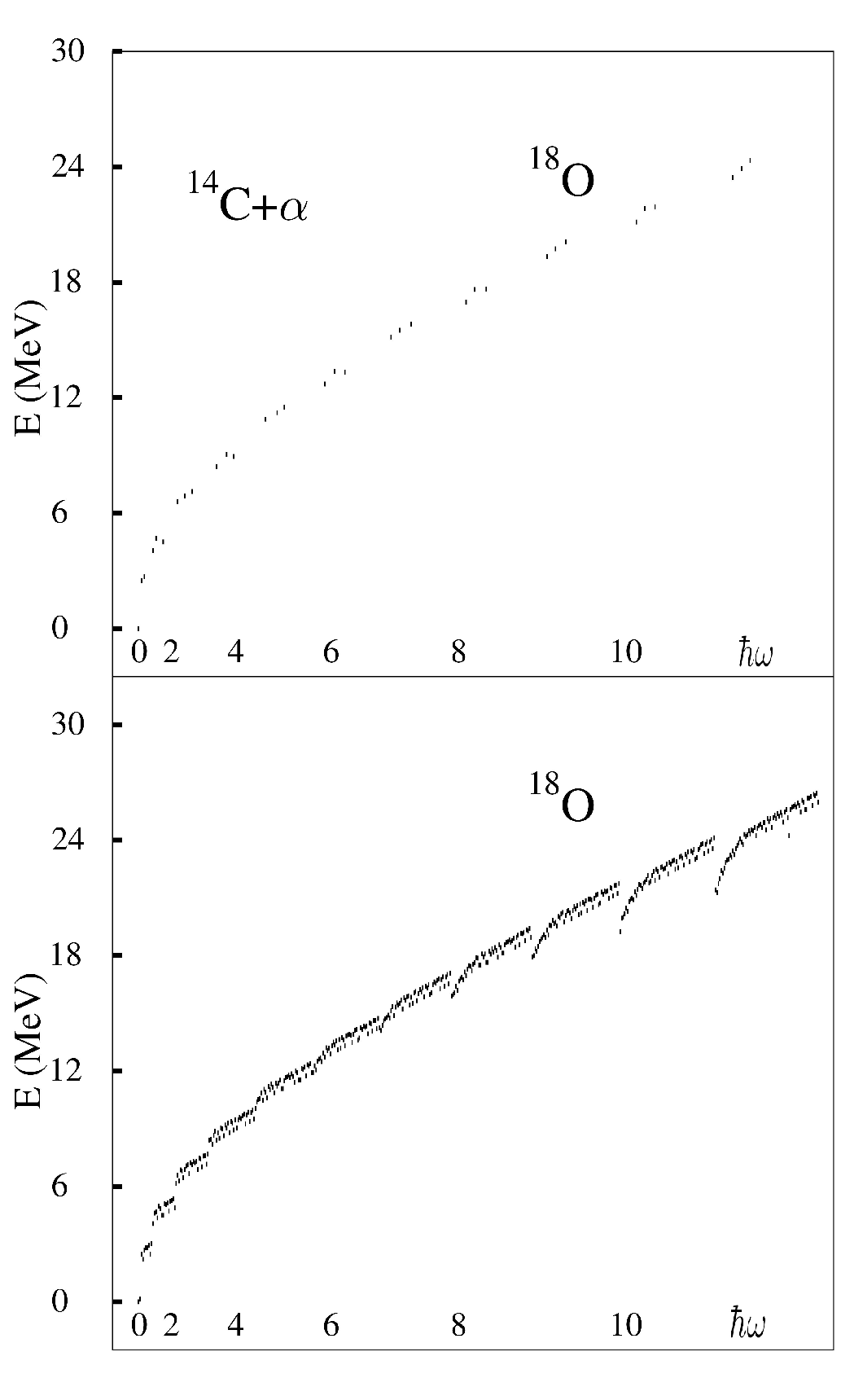}
			
			\caption{ The landscape of shell and cluster band-heads in the $^{18}$O nucleus.}
		\end{figure}

The moment of inertia of the experimental and model bands are compared in Table III.
\begin{table}[!ht]
	\centering
	\caption{The experimental values of the moment of inertia ($\theta$)  from ref.\cite{vonOertzen} and the theoretical ones obtained from equation (6). 
	}
	\begin{tabular}{|c|c||c|c|}
		\hline
		\multicolumn{2}{|c||}{Exp. Bands \cite{vonOertzen}}&\multicolumn{2}{|c|}{Th. Bands}\\
		\hline
		Bands $(K^\pi)$& $\theta$($\hbar^{2}$/MeV)& n($\lambda$,$\mu$) &$\theta$($\hbar^{2}$/MeV)\\
		\hline
		$0^{+}_{1}$&3.00&0(4,0)&1.93\\
		$1^{-}_{1}$&1.41&1(6,1)&2.13\\
		$0^{+}_{2}$&2.55&2(8,2)&2.39\\
		$0^{-}_{2}$&2.52&3(9,2)&2.48\\
		$0^{+}_{4}$&4.38&4(10,2)&2.58\\
		$0^{-}_{4}$&4.34&5(11,2)&2.69\\
		&&1(5,0)&2.00\\
		&&3(8,1)&2.30\\
		&&3(7,0)&2.15\\
		&&5(9,0)&2.31\\
		&&7(12,1)&2.68\\
		
		\hline
	\end{tabular}
\end{table}		

With the Hamiltonian obtained from the fit to the experimental band structure we can calculate the $^{14}$C+$\alpha$ cluster model spectrum and compare it to that of the alpha-scattering experiment. These resonances are not organized into bands, therefore, one can investigate the distribution of states with specific spin-parity. This is shown in Figure 6 and Table IV. We have shown the model states in the energy window of the experimental observation, defined by 0.5 MeV below the lowest-lying and 0.5 MeV above the highest-lying resonance.

				\begin{figure}[!ht]
				
				\includegraphics[width=7.0cm]{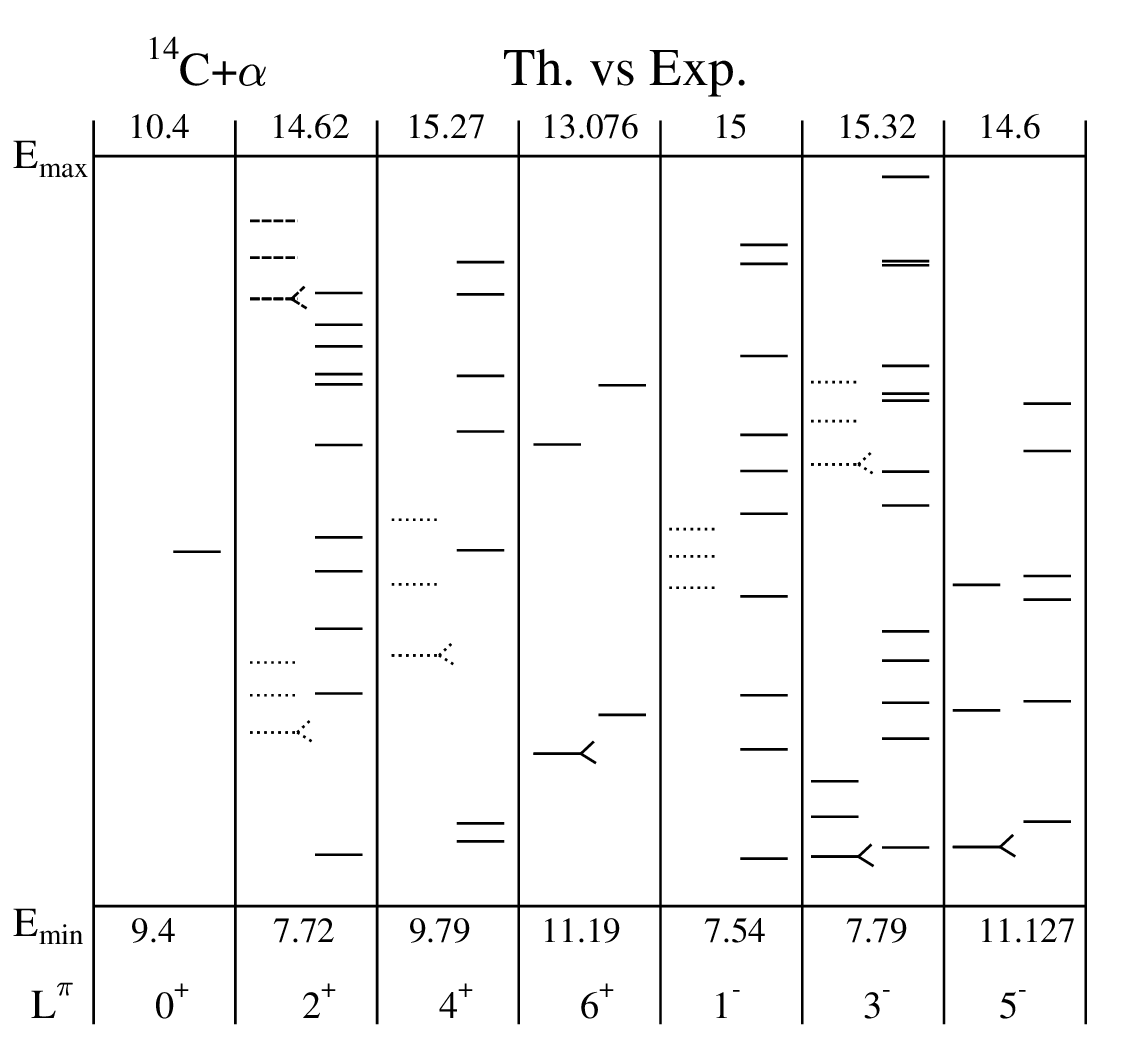}
				
				\caption{ The model spectra of the of $^{14}C+\alpha$ clusterization compared to experimental data \cite{roga2}. 
Each column shows the experimental state on the right and the model states on the left.    
Branching means degeneracy according to $K$. For positive parity states, the solid lines indicate the 2$\hbar\omega$ model space, the dotted lines represent the 4$\hbar\omega$ model space, and the dashed line the 6$\hbar\omega$ model space states. For negative parity states, the solid lines represent the 3$\hbar\omega$ model space, and the dotted lines the 5$\hbar\omega$ model space states.}
			\end{figure}
			
				
				
		\begin{table}[!ht]
			\centering
			\caption{Number of $^{14}C+\alpha$ resonance states in energy
				windows indicated by the experimental observations and model calculation \cite{roga2}. 
				}
			\begin{tabular}{ccccccccccccccccccccccccccccc}
				\hline
				\hline
				$L^\pi$&&&&$0^+$&&&&$2^+$&&&&$4^+$&&&&$6^+$&&&&$1^-$&&&&$3^-$&&&&$5^-$\\ 
				\hline
				$N_{exp}$&&&&1&&&&11&&&&7&&&&2&&&&10&&&&13&&&&6\\
				$N_{mod}$&&&&0&&&&8&&&&4&&&&3&&&&3&&&&8&&&&4\\
				\hline
				\hline
			\end{tabular}
		\end{table}

\subsection{Electromagnetic transitions}
		The B(E2) value for the intra-band transition from $ L_i$ to $ L_f$ is given by 
\begin{eqnarray}
				&B(E2, L_i\rightarrow L_f) =
				{2L_f+1\over{2L_i+1}}\alpha^2 
				\nonumber \\
				&|\left\langle(\lambda \mu)K L_i,(11)2||(\lambda \mu)KL_f\right\rangle |^2 C_{SU(3)}^{(2)}
			\end {eqnarray}
where $ \left\langle(\lambda \mu)K L_i,(11)2||(\lambda \mu)KL_f\right\rangle$ is a $SU(3)\supset SO(3)$ Wigner coefficient 
\cite{su3wigner}
and $\alpha^2(= 0.593)$ is determined by fitting the experimental value of the $2^+_1\rightarrow0^+$ transition of 3.32 W.u.; taken from NNDC \cite{compilation}. 
The transition strengths are shown in the figure 4 of the energy spectrum.
The transition operator of the above equation is that of the quadrupole operator, therefore, the inter-band transition rate is zero. (They can appear due to  symmetry breaking interactions,   or higher order transition operators.) 

\subsection{Spectroscopic factors}
				
Here we recall the basic concepts and formulas for the calculation of the alpha spectroscopic factor based on the article
\cite{Draayer}. 
We note here that the present calculation is a microscopic one, it contains no phenomenological parameters. This is a new feature in the application of the semimicroscopic algebraic cluster model, compared to previous treatments of the cluster spectroscopic factor either in a completely phenomenologic
\cite{sacm_pheno},
or in a semi-phenomenologic
\cite{sacm_semi}
manner.

\begin{table*}[!ht]
	\centering
	\caption {$[A_{nl}^{R}(^{18}O\leftrightarrows^{14}C+\alpha)]^{2}$ corresponding to different states of $^{18}$O.\label{var_table}}
	
	
	\begin{tabular}{|c|c|c|c|ccccccccc|c|}
		\cline{1-13}
		\multicolumn{2}{|c|}{$^{14}$C} & \multicolumn{11}{c|}{$^{18}$O} & \multicolumn{1}{c}{} \\
		\cline{1-13}
		\multirow{2}{*}{$(\lambda,\mu)\kappa_{L}$} & \multirow{2}{*}{$L$} & \multirow{2}{*}{n$\hbar$$\omega$} & \multirow{2}{*}{$(\lambda',\mu')\kappa_{L'}$} & \multicolumn{9}{c|}{$L'$} &  \multicolumn{1}{c}{}\\
		\cline{5-13}
		& & & & 0 & 2 & 4 & 6 & 8 & 10 & & & & \multicolumn{1}{c}{}\\
		\cline{1-14}
		 &  & 0$\hbar$$\omega$ & (4, 0)0 & 0.250 & 0.214 & 0.131 & & & & & & &\multicolumn{1}{c|}{\multirow{2}{*}{\rotatebox{-90}{$[A^{R}_{nl}]^{2}$}}}\\
		\cline{3-13}
		(0, 2)0 & 0 & 2$\hbar$$\omega$ & (8, 2)0 & 0.733 & 0.715 & 0.666 & 0.574 & 0.396 & & & & & \multicolumn{1}{c|}{\multirow{16}{*}{\rotatebox{-90}{$[A^{R}_{nl}]^{2}$}}}\\
	
		\cline{3-13}
		 &  & 4$\hbar$$\omega$ & (10, 2)0 & 0.722 & 0.709 & 0.677 & 0.618 & 0.519 & 0.345 & & & & \\

		\hline
		&&&\multirow{2}{*} {} & \multicolumn{9}{c|}{$L'$}&\multicolumn{1}{c}{}\\
		\cline{5-13}
		& & & & 1 & 3 & 5 &7 &9&&&&&\multicolumn{1}{c}{} \\
		\cline{5-14}
		&&1$\hbar$$\omega$ &(6, 1)1 &0.017 & 0.100 & 0.250 & & &&&&& \\

		&&&(5, 0)0 &0.250 & 0.204 & 0.121 & & &&&&&\multicolumn{1}{c|}{}\\

		\cline{3-13}
		&&3$\hbar$$\omega$ &(9, 2)0 &0.722 & 0.696 & 0.642 & 0.545 & 0.368 &&&&&\multicolumn{1}{c|}{}\\

		(0, 2)0&0&&(8, 1)1 &0.011 & 0.067 & 0.167 & 0.311 & &&&&&\multicolumn{1}{c|}{}\\

		&&&(7, 0)0 &0.267 & 0.236 & 0.182 & 0.103 & &&&&&\multicolumn{1}{c|}{}\\
		
		\cline{3-13}
		&&5$\hbar$$\omega$ & (11, 2)0 &0.714 & 0.695 & 0.658 & 0.596 & 0.495 &&&&&\multicolumn{1}{c|}{}\\

		&&&(9, 0)0 &0.278 & 0.256 & 0.218 & 0.162 & 0.090 &&&&&\multicolumn{1}{c|}{}\\

		\cline{3-13}
		&&	7$\hbar$$\omega$ &(12, 1)1 &0.006 & 0.036 & 0.089 & 0.167 & 0.268 &&&&&\multicolumn{1}{c|}{}\\

		\hline
		
	\end{tabular}
	
	
\end{table*}

The spectroscopic amplitude $A_{nl}(y'\longrightarrow y+x)$ for $y'$ dissociating into $y+x$ with relative motion $\phi_{nlm}(r_{xy})$
is defined in terms of the wave functions of the internal coordinates $\phi(\xi)$ of the corresponding nuclei
	\begin{equation}
		A_{nl}(y',y)=\begin{pmatrix} y' \\ y \end{pmatrix}^\frac{1}{2} \int[\phi_{y}(\xi_{y})\phi_{x}(\xi_{x})\phi_{nlm}(r_{xy})]^{\ast}\phi_{y'}(\xi_{y'})d\xi_{y'}.
	\end{equation}
Here the binomial coefficient expresses the consequences of antisymmeterization \cite{Austern}.

The spectroscopic strength for the $A(a,b)B$,  $a=b+x$, $B=A+x$ reaction is 
		\begin{equation}
		A_{nl}^{n'l'}=\sum A_{nl} (B\rightarrow A+x)  A_{n'l'}^*(a \rightarrow b+x). 
	\end{equation}
The summation  is over all possible intermediate states formed by the $x$ transferred nucleons.
If the $x$ particles are transferred in a single pure configuration
(e.g., $(0s)^{4}$ for an $\alpha$-particle),  the sum reduces to one term.  Then the structure factor is equal to the product of two spectroscopic amplitudes, one for the $B  \rightarrow A + x$ vertex and the other for the $a \rightarrow b + x$ vertex.

	For eigenstates of the harmonic oscillator, Elliott and Skyrme \cite{Elliot} established a simple relation between wave functions described in terms of internal and space fixed coordinates ($\phi$ and $\psi$ respectively). Using their results, the spectroscopic amplitude can be written in the form  
	\begin{equation}
		\begin{split}&A_{nl}(y'\rightarrow y+x)=\binom{y'}{y}^{\frac{1}{2}}(\frac{y'}{y})^{\frac{1}{2}n}\\&\sum \left\langle\psi_{y};\psi_{x}|\}\psi_{y'}\right\rangle	\left\langle\phi_{x}(\xi_{x})\phi_{nlm}(r)|\psi_{x}\right\rangle \end{split}
	\end{equation}	
where each $\left\langle\psi_{y};\psi_{x}|\}\psi_{y'}\right\rangle$ is a shell model coefficient of fractional parentage (c.f.p) for the separation of $x$ nucleons in an eigenstate $|\psi_{x}\rangle$ from the state $|\psi_{y'}\rangle$ leaving the $y=y'-x$ nucleons in the state $|\psi_{y}\rangle$. The factor $\left\langle\phi_{x}(\xi_{x})\phi_{nlm}(r)|\psi_{x}\right\rangle$, (the “G” of ref. \cite{Ichimura}), is simply the overlap of the $x$ separated nucleons in a  state $|\psi_{x}\rangle$  with x nucleons having an internal wave function $\phi_{x}(\xi_{x})$ and a motion relative to a space fixed coordinate system given by $\phi_{nlm}(r)$. The summation in equation (19) is over all possible $|\psi_{x} \rangle$ consistent with both $|\psi_{y} \rangle$, $|\psi_{y'} \rangle$ and  $\phi_{x}(\xi_{x})$, $\phi_{nlm}(r)$. Within a single major oscillator shell the sum can be reduced to a single term if the transferred nucleons are in a simple $(0s)^{x}$ configuration with internal SU(3) symmetry (0,0).

Now we are going to make use the SU(3) $LS$ coupled formalism of the shell model.
Let us suppose that  both the initial and final states of the target and residual nuclei have pure $[f]\alpha(\lambda\mu)\beta ST$ symmetry labels, where 
$[f]$ denotes the space symmetry of the major shell,
$(\lambda\mu)$ are the SU(3) quantum numbers,
$\alpha$ and $\beta$ are multiplicity labels, and
$ST$ stand for the spin and isospin.
Then the general result which involves a sum over all the $[f]\alpha(\lambda\mu)\kappa_{L}L\beta ST$ quantum numbers reduces to 
		  \begin{equation}
		  	\begin{split}
		  		A_{nlsj}(y'\rightarrow y+x)  = G\binom{y'}{y}^{\frac{1}{2}}\left(\frac{y'}{y}\right)^{\frac{1}{2}n} \\
		  		\times \left\langle [f]\alpha(\lambda\mu)\beta ST;[x](n0)lst|\}[f']\alpha'(\lambda'\mu')\beta'S'T'\right\rangle \\
		  		\times \left\langle TM_{T};t m_{t}|T'M'_{T}\right\rangle \\
		  		\times \sum_{\kappa_{L}L\kappa_{L'}L'} cc'X\begin{Bmatrix}L &S& J\\ l& s& j\\ L'& S' &J'\end{Bmatrix} \\
		  		\times \left\langle (\lambda\mu)\kappa_{L}L;(n0)l||(\lambda'\mu')\kappa'_{L}L'\right\rangle .
		  	\end{split}
		  \end{equation}

Here 
$X\{\}$ is a unitary 9j-symbol which accounts for the conversion from jj to LS coupling \cite{A. R. Edmonds}. $\left\langle (\lambda\mu)\kappa_{L}L;(n0)l||(\lambda'\mu')\kappa'_{L}L'\right\rangle$ is a reduced SU(3)$\supset$R(3) Wigner coefficient \cite{J. D. Vergados}, and c and c' are the  expansion coefficients for the wave functions of the target and residual nuclei in terms of the SU(3)-LS coupled basis states. As in ref. \cite{Ichimura}, G is used to represent the overlap integral 
$\left\langle\phi_{x}(\xi_{x})\phi_{nlm}(r)|\psi_{x}\right\rangle$.
The factor
$\left\langle [f]\alpha(\lambda\mu)\beta ST;[x](n0)lst|\}[f']\alpha'(\lambda'\mu')\beta'S'T'\right\rangle$
is an $x$-particle 
c.f.p. \cite {Akiyama}.

	
	A reduced (or relative) spectroscopic amplitude is defined 
as the result of the multiplication of the last three factors of (20).
In what follows we are going to calculate these factors for the SU(3) basis states, therefore the $c$ and $c^{\prime}$  coefficients are 1.
	\begin{equation}
		\begin{split}
			A_{nlsj}^R(y' \rightleftarrows y+x)
			=\sum_{\kappa_{L}L\kappa_{L'}L'}C_{\kappa_{L}L}C'_{\kappa_{L'}L'}\\
			\times X\begin{Bmatrix}L &S& J\\l & s& j\\ L'& S' &J'\end{Bmatrix}\\
			\times	\left\langle (\lambda\mu)\kappa_{L}L;(n0)l||(\lambda'\mu')\kappa'_{L}L'\right\rangle \end{split}.
	\end{equation}
$A^R_{nlsj}$  carry the same spin dependence as the $A_{nlsj}$ of (20) and hence, relative to a particular transfer (e.g., ground state to ground state), the two are identical.\\
Since spectroscopic factors are proportional to the square of the corresponding amplitude, we tabulated in Table V
the square of the amplitudes.

\section{Summary and conclusions}

We have applied a symmetry-adapted model for the description of the spectrum of the $^{18}$O nucleus. In particular, the multiconfigurational dynamical symmetry was used, which represents the common intersection of the shell, collective, and cluster models for the multi-shell problem. The basis states are characterized in each configuration by the 
$\rm U_s(3) \otimes U_e(3) \supset U(3) \supset SU(3) \supset SO(3) $
symmetries, and a Hamiltonian was used, which is invariant with respect to the transformations between the different configurations (containing only harmonic oscillator, quadrupole and $L \times L$ interaction). Therefore, all three configurations, proposed earlier,  i.e. the shell, 
$^{14}$C+$^{4}$He and $^{12}$C+$2n$ +$^{4}$He are present in each state of the model spectrum. 

The band structure, established in
\cite{vonOertzen}
and the high-lying states from the alpha-scattering
\cite{roga2}
was described in a unified way. The lowest-lying three bands have states in which the overlap of the three relevant configurations is 100\%. In addition to the ground state another shape isomer of $2 \hbar \omega$ excitation is contributing to the core-plus-alpha spectrum. A prolate superdeformed shape is also found with [12,4,2](8,2) quantum numbers, that seems to correspond to the [12,4,0](8,4) SD state of $^{16}$O
\cite{lepcsok}.  Some of the high-lying states found in alpha-scattering may sit in this third minimum of the energy surface. The relative alpha spectroscopic factors have been determined, too.

It seems that the $^{14}$C+$^{4}$He  cluster spectrum with a surprisingly simple dynamically symmetric Hamiltonian is able to account for the gross features of the observed energy spectrum of the $^{18}$O nucleus.  
The energy range spans from the ground state to high-lying excitations, 
and the $^{14}$C+$^{4}$He  states have an overlap with the shell model as well as with the $^{12}$C+$2n$+$^{4}$He configurations.

{}

\end{document}